# Structure Profile of Complex Networks by a Model of Precipitation

Zhenggang Wang and K.Y. Szeto*

Department of Physics,

Hong Kong University of Science and Technology,

Clear Water Bay, Hong Kong SAR

* Corresponding author: phszeto@ust.hk

**Abstract**

The organizational structure of a network is investigated with a simulated precipitation model which does not make use of prior knowledge about the community structure of the network. The result is presented as a structure profile through which various definitions of communities can be applied for specific applications. The simulated precipitation model performs the grouping of nodes so that nodes belonging to the same "community" automatically aggregate, thereby revealing regions of the adjacency matrix with denser interconnections. The process is analogous to massive particles precipitating towards the lower potential layer. Without loss of the infrastructure information, a community structure profile of a network can be obtained as the ground state of the Hamiltonian. The method is also applicable to directed and weighted networks.

**I. Introduction**

Numerous complex systems, such as the biological systems[1-4], social networks[2, 5-10] and the information systems[11], can be studied in the form of networks, with nodes (objects under study) and edges describing the interactions between nodes. According to the types of edges, there are four basic types of networks, weighted (valued, positive or negative), un-weighted (binary), directed and undirected, with the weighted directed network as the most general one. An analysis of the topology of networks provides important clues to the intrinsic properties of the system [12-15]. Recently, many algorithms have been invented to address the topological properties, especially in the hidden structures corresponding to "communities" [8, 12, 16-22]. These algorithms of community detection basically aim at profiling the organizational structure of the network by finding groups of nodes with high density of interconnections. For the communities identified, their sizes, positions, associated members and edge density are recorded, but usually it is difficult to reveal hidden relations on the infrastructures, not to mention the complex nature of the overlapping or hierarchical structures. Consequently, the structure profiles of the network revealed in these community detection methods are presented as squares along the diagonal of the adjacency matrix, with the resolution limit given by the smallest size of communities identified [23]. One usual requirement on these methods is to assume certain prior knowledge about the network, such as the critical size, density or the boundary of the communities. The performance of these algorithms quickly deteriorate when the community-like structure is fuzzy [23]. Furthermore, most of these methods are difficult to use in the case of weighted or directed networks.

Our method of community detection is based on a model of precipitation of the entries of the adjacency matrix in the form of a structure profile. The results have three key advantages over existing methods: (1) applicable to directed and weighted networks, (2) no prior knowledge of the community size, density parameter, etc are needed, and (3) flexible in different applications for those who aim at specific details of the organizational structure of the network. One last advantage which requires further investigation is that our algorithm reveals new information on the infrastructure, with the possibility of gaining some insights into the growth mechanism of the network.

To illustrate our method, let's consider a directed weighted network with its interaction matrix (adjacency matrix) shown in Fig.1. In this network, the attachment preference is not uniform between nodes, so the shapes of the interaction community structures are layers of annulus and not squares. Inside the outer-most ring (community), there is a sub-community structure with higher weights. There are five ring-like communities with weight 1-5 from outer to the center. The stylish pattern presented in the matrix provides us a lot of information, such as the attachment preference, hierarchical community structures and so on. For example, for nodes with labels near *N/2(=50* in Fig.1), we can say that they have links to many other nodes, for the density of links along the middle rows are high in the disk like structure profile in Fig.1. For nodes with labels far from the middle, such as node number 20 and 80, there are fewer links with other nodes, although they are still linked to node number 50. The information provided on the infrastructure of this network is hard to observe from the block diagrams of the community aligned along the diagonal, but can be seen easily from our ring like structure profile in Fig.1. Indeed, if we use the usual community detection methods, it is difficult to use 'squares' (communities identified) to piece together the original structures (rings).

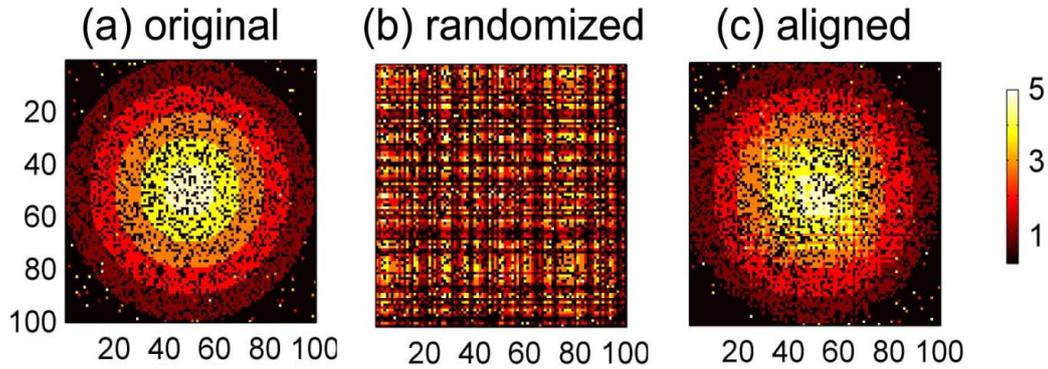

FIG. 1. Identification of complex community structures in weighted directed networks: (a) an artificial network with complex community structures, which is a series of ring-like community structures with different weights. In the rings, the density of edges is uniform, 0.8, but the weights are different and accordingly colored (1-5, shown as the color bar). (b) the adjacency matrix after randomizing the node labels in (a). (c) a configuration of the aligned matrix reconstructed from (Fig. 1b) with our method.

In our method, instead of picking out the communities directly, we present the boundaries of communities to provide a global profile of the network structure. As a guiding principle in drawing these boundaries, regions in the adjacency matrix with high density of edges (weighted or un-weighted), are considered as the core structures and identified with priority. Our method is analogous to the description of the precipitation of small impurities in a stirred liquid. After the liquid is thoroughly stirred, the impurities start to precipitate due to the gravity. The system achieves equilibrium at the state with lowest gravitational potential. The impurities with higher density rest in the lower layer.

For a general network, we can describe the interaction between nodes by an adjacency matrix $w$, though the ordering of nodes is arbitrary. The usual problem of community detection is a renaming scheme of these nodes so as to reveal the hidden topological relations between nodes. For a simple un-weighted undirected network, $w$ is a symmetric matrix and its entries $w_{ij}$ is either 1 or 0, with 1 denoting the existence of a link between node *i* and node *j* and 0 for no connection. In our model, we consider the link as granular particle with certain mass proportional to the value of $w_{ij}$. We introduce a gravitational field that is perpendicular to the diagonal of the adjacency matrix, and symmetric on both sides. We then re-align the matrix ordering to facilitate the links to 'precipitate' towards the diagonal. In this precipitation process, the movement of the links is realized by relabeling (or re-ordering) of the nodes of the adjacency matrix that does not alter the topology of the network. To reach a lower energy state, the community structures with higher link density (or mass density) will automatically aggregate closer to the diagonal (the 'ground level'). In this precipitation process, the total energy of the system decreases. At equilibrium, the ground state of the Hamiltonian is reached and the nodes belonging to the same "community" will be grouped together. Eventually, the adjacency matrix will be aligned, resulting in a high resolution profile of the community structures. This process is shown in Fig.1 with (a) as the original structure, (b) the randomized adjacency matrix, and (c) the result of applying our precipitation process.

To discuss the detail of our model, let's introduce the Hamiltonian of the system. Given a link with (*x,y*) coordinate in the adjacency matrix (Fig.2) given by (*a,b*), we can compute its perpendicular distance towards to the diagonal of the matrix as $d_{ab} = |a-b|/\sqrt{2}$. This link has a weight $w_{ab}$ as it connects node *a* and node *b*. Next we introduce a potential for a particle of mass *m* at a height $h = d_{ab}$ above ground in a gravitational field *g*. The analog for the potential *"mgh"* is then $mgh = w_{ab}d_{ab}$ with its minimum at the diagonal of the matrix (minimum $d_{ab} = 0$). By considering the links as particles with different mass, we obtain the following Hamiltonian of the system,

$$H(E) = \sum_{(i,j) \in E} w_{ij} d_{ij} = \frac{1}{\sqrt{2}} \sum_{(i,j) \in E} w_{ij} |i-j|, \quad (1)$$

where $w_{ij}$ is the weight of the link between node *i* and *j*, and *E* is the set of links in a certain configuration, i.e, in a particular node labeling corresponding to the given adjacency matrix. Since in this Hamiltonian, the distance between two nodes is measured by their ordering and the interaction valued by the weight, we can also relate *H* to the weighted distance between all the nodes,

$$L = \sum_{(i,j) \in E} w_{ij} |i-j| = \sqrt{2} H . \quad (2)$$

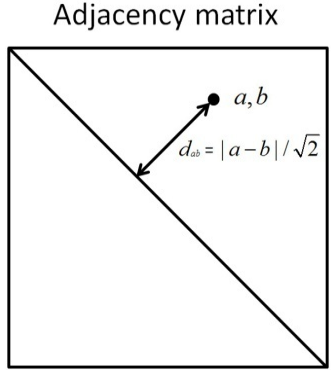

FIG. 2. The precipitation model in the *x-y* plane. The perpendicular distance of a link between two nodes (*a,b*) with respect to the diagonal of the adjacency matrix is $d_{ab} = |a-b|/\sqrt{2}$. The factor $\sqrt{2}$ comes from the fact that the adjacency matrix is a square matrix. The total 'ordering distance' is denoted as *L* in Eq.(2). Here (*a,b*) denotes the (*x,y*) coordinate of the link in the adjacency matrix. Also, (*a, b*) refers to the ordering of nodes *a* and *b*.

In the simulated precipitation process, new configuration *E'* is generated by exchanging any two nodes in a given configuration *E*. Our aim is to find the configuration that minimizes the total energy. We can achieve this by simulated annealing [24] with the following cooling schedule: the system starts from an initially high temperature and cooled down in multiple steps. At the *k*-th step, the temperature $T_k$ decreases with by a factor $\alpha = 0.99$ so that $T_{k+1} = \alpha T_k$. In each step, we ensure that we have performed exchanges between all pairs of nodes (for a network with size *N*, there are *N(N-1)/2* pairs). The stopping criterion is that the algorithm either has executed a preset maximum number of steps or when no further exchange is accepted at the given low temperature. During the cooling process, the overall effect is that the weighted links are precipitating towards to the diagonal. The movement of the link is determined by the re-alignment of the ordering of nodes, which is a one-dimensional process: all gravitating towards the diagonal. The time complexity of the algorithm depends on the specific scheme of the generation of new configuration (new ordering) and the details of the simulated annealing. In our scheme, new configurations are generated by exchanging all pairs of nodes at each step so that the time complexity of the algorithm is O(*N²*). It ensures that at each step the nodes can be completely realigned. One can improve this efficiency but here we focus on the methodology and the general description of the method. We have also studied the problem of the stability of the simulated annealing algorithm with repeated cooling schedule, in order to ensure that the final result is not a trapped local minimum of the total energy. In the repeated cooling scheme, the ordering of the nodes was randomized at high temperature, slowly condensed during cooling process and finally rested in the minimum energy state. For all the examples investigated, the same structure profile is obtained, so that the problem of being trapped at local minimum has not occurred in the precipitation process. As expected, for the networks with sparse link density, the time complexity can be improved. For a more detailed analysis on the computation cost of large networks, we propose a fast scheme which makes a compromise between efficiency and cost. In this scheme, exchange only occurs between neighboring nodes in each step during the cooling process. The time complexity becomes linear in the size of network, $O(N)$.

We now illustrate our method with several examples.

**A. Structure profile of irregular community structures**

To test our algorithm, we consider several artificial networks that are directed and weighted. An example is shown in Fig.1a. We shuffle the ordering of the nodes in Fig.1a to get our randomized initial configuration shown in Fig.1b. Without the introduction of any parameter, we obtained the aligned matrix (Fig.1c) by applying our method to Fig.1b. We can see that Fig.1c is a good approximation to the original data Fig.1a. The original profile in Fig.1 shows us the difficulty of applying the existing methods of community detection which attempts to approximate the ring with blocks of various sizes along the diagonal of the adjacency matrix. In general, it is difficult to obtain high resolution profile with existing community detection algorithms for many complex network, as the identification of communities is very sensitive to the preset criterion in the detection methods. To further illustrate this point, let's construct an artificial network based on a sequential addition of nodes, with links between any two nodes (i.e. at site $i$ and $j$) built with the probability,

$$p_{ij} = \begin{cases} q^{|i-j|}, & \text{if } i \neq j \\ 0, & \text{if } i = j \end{cases}, \quad (3)$$

here $0 < q < 1$ and no self-connection assumed (this assumption is not necessary, but simplify our discussion). In this network, the problem of using the community detection algorithms is that we have to identify the 'communities' from a set of similar groups of nodes. The number of communities identified depends on the parameters used in the definition of community, such as the preset size of the community, their numbers, or the critical density of links in a block to be considered as a community. In this artificial network, since the probability of having a link approaches 1 as we get closer to the diagonal: $p_{ij}|_{i \neq j} = q^{|i-j|} \xrightarrow[|i-j|]{} 1$, the link density $\rho$ increases as the distance $|i-j|/\sqrt{2}$ of the link from the diagonal decreases. Now, consider a set of links with an average distance $K$ from the diagonal of the adjacency matrix. The link density $\rho(K)$ for this set of links will be greater as $K$ decreases. Therefore, the probability that $\rho(K)$ is greater than a critical value $\rho_c$ will be greater as $K$ decreases. If we identify a community in this network using a critical link density $\rho_c$, we will find more communities of smaller size than with larger size. Indeed, since smaller community has a smaller $K$, thereby higher $\rho(K)$, leading to higher probability to be detected as a community, we will see a dramatically larger number of communities identified as $K$ decreases towards zero. We see from this example the difficulty of community detection with the traditional method.

Another common task in community detection is to search for all communities of size smaller than $N^*$, which is a preset parameter. In our artificial network (Eq.3), this task can be translated into one of community detection for set of nodes with link density greater than the following critical density

$$\rho_c = \frac{2}{N^{*2}} \sum_{i=1}^{N^*-1} (N^* - i) q^i . \qquad (4)$$

In this situation, we see that there are many more communities because all communities will have link density $\rho(K)$ greater than $\rho_c$ for $K$ less than $N^*$. With this example, we see the difficulty encountered in the usual community detection algorithms, as the information of the structure of the network must be described by a large number of ever smaller communities. However, with our precipitation model, the infrastructure described by the structure profile can still be very useful. Indeed, once the structure profile is known, one has a visual picture of the network structures and can pick out the communities of interest from the profile directly.

We now use our method to obtain the structure profiles on four different artificial networks: the artificial network generated by Eq.3, the Erdős-Rényi (ER) network [25], and the Barabási-Albert (BA) network [26] and the artificial network containing five overlapping ER networks. In the artificial network (Fig.3a), the high resolution profile shows increasing link density towards the diagonal. In the ER network, small communities generated by random connections are aligned along the diagonal. For the bigger community, the link density decays quickly to the average density, resulting in a profile with a smooth link distribution (evenly distributed on the adjacency matrix) and a bright line along the diagonal (Fig.3b). In the BA network, the nodes with large number of edges are squeezed to the center and the same bright line along the diagonal can be observed, similar to the ER network (Fig.3c). This method can successfully separate the five overlapping ER communities (Fig. 3d is obtained with the same algorithm as those shown in Fig. 3a-c. No additional parameter such as the number of communities is used). In each community shown in Fig.3d, the structure profile is similar to the one shown in Fig. 3b. These four networks show characteristic infrastructure which are difficult to be described by the usual block community structures.

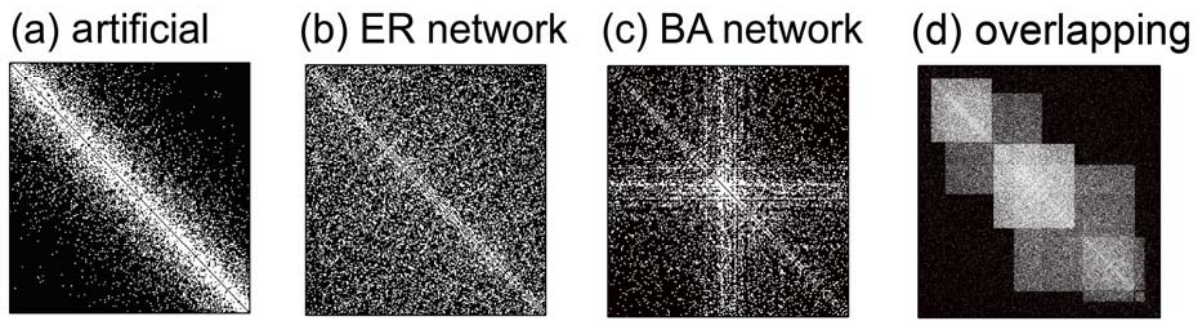

FIG.3. After being randomized, the adjacency matrix of four different networks (with the size, N=200 for (a-c) and N=800 for (d)) are reproduced from the simulated precipitation process, (a) artificial network generated by Eq. 3, (b) Erdős-Rényi (ER) network, (c) Barabasi-Albert (BA) network (the bright pixel stands for an edge) and (d) artificial network containg five overlapping communities. (a) The artificial network is generated with parameter $q=0.96$, defined in Eq.3. (b) The ER network is generated with the probability of connection between any two nodes, $p=0.3$. The aligned matrix is mostly uniform except for some small communities generated by random connections, which are aligned along the diagonal. (c) The BA network is generated with the initial number of nodes, $m_0=3$, and the average number of links, $m=30$. The nodes with large number of edges are centered and the small communities formed by random connection are aligned along the diagonal. (d) An artificial network contains five overlapping communities, which are generated by the ER model with the different edge densities. We first randomize the node labels to get something similar to Fig.1b and then apply

our algorithm to get the structure profile. Without requiring the application of conventional method of community detection or additional parameters, the structure profile already reveals the five overlapping communities.

**B. Dolphin, Zachy Karate club and the college football network**

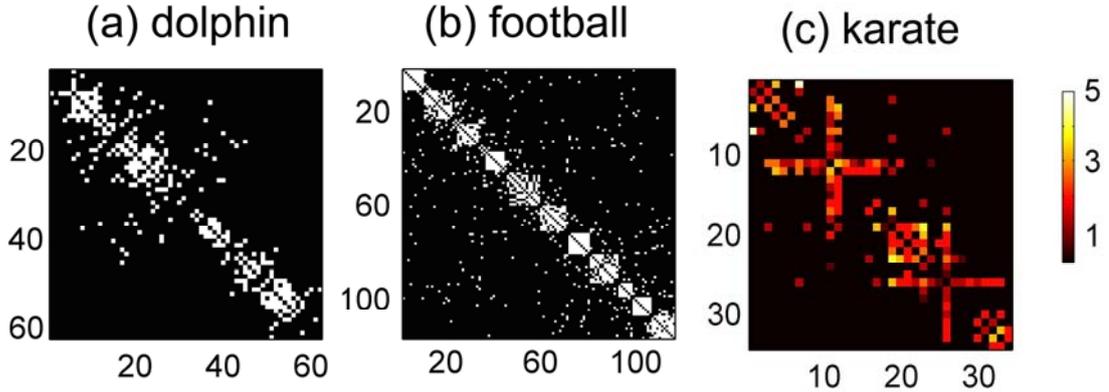

FIG.4. The structure profile of three benchmark networks (a) the bottlenose dolphins of Doubtful Sound [27] (b) the college football network (binary) [28] and (c) Zachy karate club network (weighted) [29].

We consider three benchmark networks which structures have been investigated by many methods [17, 18]. The dolphin network contains 62 dolphins living in Doubtful Sound, New Zealand and the ties are recorded according to the observation of statistically significant frequent association (Fig. 4a).. The college football network [28] represents the game schedule of the 2000 season of Division I of the U.S. college football leagues (Fig. 4b). The Zachy karate club network shows the pattern of friendships between the members of a karate club at an American university in the 1970s [29] and the entries are presented as the presence or absence of ties among the members of the club and weighted by the number of situations in and outside the club in which interactions occurred (Fig. 4c).. The profiles reveal the similar structures/communities presented in Ref.[17, 18, 22, 28] and the infrastructures of the communities (sub-communities) are revealed also by our method.

**C. Bernard & Killworth HAM radio network**

We here use our algorithm on a weighted undirected network, presented as a 44×44 matrix, which is a social network recording the interactions between the members of the HAM radio amateurs over a month by Bernard and Killworth [31]. The weight is the number of contacts between each pair of members. The structure profile of the network has a butterfly-like pattern in its adjacency matrix showing complex infrastructures. When we apply our precipitation model to this data, we obtain an adjacency matrix shown in Fig.5, where we see a clique (the 5×5 white square aligned in the center of the graph) of five members who maintains frequent contact with all the members in the clique. These five members also keep frequent contact with the other members, as can be revealed from the bright cross pattern shown in Fig.3. Besides these two observations, the structure profile obtained with our method does reveal rich details on the infrastructures which provide direct material for the further study of the social interactions in this network.

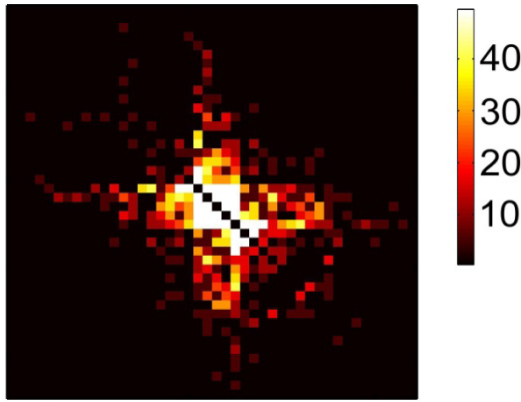

FIG.5. The structure profile of the Bernard & Killworth HAM Radio network, which is a weighted undirected network. It records the amateur HAM radio calls made over a one-month period, as monitored by a voice-activiated recording device.

**D. Structure profile of Slovenian Journals network**

The Slovenian Journals network is a weighted undirected network, with size 124×124. In 1999 and 2000, survey was conducted over 100,000 people to collect the journals and magazines read (source CATI Center Ljubljana [32]). There are a total of 124 different journals and magazines listed. The results are summarized in a weighted network, with the weight of element (A, B) denoting the number of people who have read both journal A and journal B. The elements on the diagonal denote the number of people who have read a particular journal. The result of profiling the structure is shown in Fig.6. We observe a bright cross at the center, with some overlapping bright community and some infrastructures in these bright squares. These details are difficult to find with existing methods of community detection which restrict a profiling with squares.

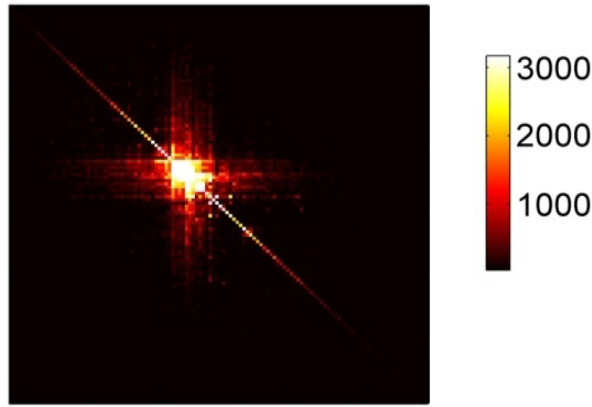

FIG.6. The structure profile of the Slovenian Journals weighted undirected network.

In conclusion, we present a simple method for profiling the infrastructure of a complex networks, which not only contains the usual block structures of the community, but various complex organizational infrastructure of the system. We emphasize that our simulated precipitation model provides a new perspective in analysing network topology by profiling its structure with general applicability to directed and weighted networks. In the precipitation process, we make use of a specific Hamiltonian, but the idea of minimizing energy for the precipitation process is general, and the Hamiltonian used only serves as a simple example. One may use other Hamiltonian for the minimization process to improve the efficiency for certain special case, but that is of secondary importance to our thesis. Our method avoids using prior knowledge of the network, for instance, the size, the critical density and various features of members, which may depend on the interests of the investigator. For general networks, our method can pre-process the data before the application of conventional techniques that take advantage of the prior information, such as the known number of communities. However, a more important application of our method is on those data without prior information of the community. Our method can reveal the community structure without pre-setting parameters such as the size and number of communities. We will discuss the ordering distance in future work, as this distance is a valuable quantity depicting the relationship between the nodes in a network. We have shown that the time complexity in general is $O(N^2)$ and can be reduced to $O(N)$. Further improvement can be made with special optimization techniques. Finally, we like to summarize by restating that our algorithm has three important advantages: no need to set community structure parameters in advance, the full information of the infrastructures, and applicability to all kinds of networks (directed, undirected, weighted or un-weighted).

**Acknowledgements**    K.Y. Szeto acknowledges the support of CERG grant 602506.

32. http://vlado.fmf.uni-lj.si/pub/networks/data/2mode/journals.htm